\documentclass{article}
\usepackage{amssymb}
\usepackage{amsmath}

\input{tcilatex}

\begin{document}

\title{Varying Alpha}
\author{John D. Barrow \\
DAMTP\\
Centre for Mathematical Sciences\\
Cambridge University\\
Cambridge CB3 0WA\\
UK}
\maketitle

\begin{abstract}
We review properties of cosmological theories for the variation of the fine
structure 'constant'. We highlight some general features of the cosmological
models that exist in these theories with reference to recent quasar data
that are consistent with time-variation in the fine structure 'constant'
since a redshift of 3.5.
\end{abstract}

\section{ Introduction}

There are several reasons why the possibility of varying constants should be
taken seriously \cite{con}. First, we know that the best candidates for
unification of the forces of nature in a quantum gravitational environment
only seem to exist in finite form if there are many more dimensions of space
than the three that we are familiar with. This means that the true constants
of nature are defined in higher dimensions and the three-dimensional
shadows\ we observe are no longer fundamental and need not be constant. Any
slow change in the scale of the extra dimensions would be revealed by
measurable changes in our three-dimensional 'constants'. Second, we
appreciate that some apparent constant might be determined partially or
completely by spontaneous symmetry-breaking processes in the very early
universe. This introduces an irreducibly random element into the values of
those constants. They may be different in different parts of the universe.
The most dramatic manifestation of this process is provided by the chaotic
and eternal inflationary universe scenarios where symmetries determining
both the number and the strength of forces in the universe at low energy can
break differently in different regions. Third, any outcome of a theory of
quantum gravity will be intrinsically probabilistic. It is often imagined
that the probability distributions for observables will be very sharply
peaked but this may not be the case for all possibilities. Fourth, a
non-uniqueness of the vacuum state for the universe would allow other
numerical combinations of the constants to have occurred in different
places. String theory indicates that there is a huge 'landscape' ($>10^{500}$%
) of possible vacuum states that the universe can find itself residing in as
it expand and cools. Each will have different constants and associated
forces and symmetries. It is sobering to remember that at present we have no
idea why any of the constants of Nature take the numerical values they do
and we have never successfully predicted the value of any dimensionless
constant in advance of its measurement. However, the last reason to consider
varying constants is currently the most compelling. For the first time there
is a body of detailed astronomical evidence for the time variation of a
traditional constant. The observational programme of Webb et al \cite%
{webb1,webb2} has completed detailed analyses of three separate quasar
absorption line data sets taken at Keck and finds persistent evidence
consistent with the fine structure constant, $\alpha $, having been \textit{%
smaller} in the past, at $z=1-3.5.$ The shift in the value of $\alpha $ for
all the data sets is given provisionally by $\Delta \alpha /\alpha
=(-0.57\pm $ $0.10)\times 10^{-5}.$ This result is currently the subject of
detailed analysis and reanalysis by the observers in order to search for
possible systematic biases in the astrophysical environment or in the
laboratory determinations of the spectral lines. So far it has not been
undermined or confirmed by other observations (for the most reason
discussion of the status of uncertainties, see \cite{webb3}).

The first investigations of time-varying constants were those made by Lord
Kelvin and others interested in possible time-variation of the speed of
light at the end of the nineteenth century. In 1935 Milne devised a theory
of gravity, of a form that we would now term 'bimetric', in which there were
two times -- one ($t$) for atomic phenomena, one ($\tau $) for gravitational
phenomena -- linked by $\tau =\log (t/t_{0})$. Milne \cite{mil} required
that the 'mass of the universe' (what we would now call the mass inside the
particle horizon $M\ \approx c^{3}G^{-1}t$) be constant. This required $%
G\varpropto t.$ Interestingly, in 1937 the biologist J.B.S. Haldane took a
strong interest in this theory and wrote several papers \cite{hal} exploring
its consequences for the evolution of life. The argued that biochemical
activation energies might appear constant on the $t$ timescale yet increase
on the $\tau $ timescale, giving rise to a non-uniformity in the
evolutionary process. Also at this time there was widespread familiarity
with the mysterious 'large numbers' $O(10^{40})$ and $O(10^{80})$ through
the work of Eddington (although they had first been noticed by Weyl \cite%
{weyl} -- see ref. \cite{bt} and \cite{con} for the history). These two
ingredients were merged by Dirac in 1937 in a famous development (supposedly
written on his honeymoon) that proposed that these large numbers $(10^{40})$
were actually equal, up to small dimensionless factors. Thus, if we form $%
N\sim c^{3}t/Gm_{n}\sim 10^{80}$, the number of nucleons in the visible
universe, and equate it to the square of $N_{1}\sim e^{2}/Gm_{n}^{2}\sim
10^{40},$ the ratio of the electrostatic and gravitational forces between
two protons then we are led to conclude that one of the constants, $%
e,G,c,h,m_{n}$ must vary with time. Dirac \cite{Dir} chose $G\varpropto
t^{-1}$ to carry the time variation. Unfortunately, this hypothesis did not
survive very long. Edward Teller \cite{tell} pointed out that such a steep
increase in $G$ to the past led to huge increases in the Earth's surface
temperature in the past. The luminosity of the sun varies as $L\varpropto
G^{7}$ and the radius of the Earth's orbit as $R\varpropto G^{-1}$ so the
Earth's surface temperature $T_{\oplus }$ varies as $(L/R^{2})^{1/4}%
\varpropto G^{9/4}\varpropto t^{-9/4}$ and would exceed the boiling point of
water in the pre-Cambrian era. Life would be eliminated. Gamow subsequently
suggested that the time variation needed to reconcile the large number
coincidences be carried by $e$ rather than $G$, but again this strong
variation was soon shown to be in conflict with geophysical and radioactive
decay data. This chapter was brought to an end by Dicke \cite{dicke} who
pointed out that the $N\sim N_{1}^{2}$ large number coincidence was just the
statement that $t$, the present age of the universe when our observations
are being made, is of order the main-sequence stellar lifetime, $t_{ms}\sim
(Gm_{n}^{2}/hc)^{-1}h/m_{n}c^{2}\sim 10^{10}$ yrs, and therefore inevitable
for observers made out of chemical elements heavier than hydrogen and
helium. Dirac never accepted this anthropic explanation for the large number
coincidences (believing that 'observers' would be present in the universe
long after the stars had died) but curiously can be found making exactly the
same type of anthropic argument to defend his own varying $G$ theory by
highly improbable arguments (that the Sun accretes material periodically
during its orbit of the galaxy and this extra material cancels out the
effects of overheating in the past) in unpublished correspondence with Gamow
in 1967 (see \cite{con} for fuller details). Dirac's biographer has revealed
that in 1993 he expressed 'an article of faith...\ that the human race will
continue to live for ever and will develop and progress without limit' \cite%
{farm}. This belief motivates his comments relating to the anthropic
argument.

Dirac's proposal acted as a stimulus to theorists, like Jordan, Brans and
Dicke \cite{bd}, to develop rigorous theories which included the time
variation of $G$ self-consistently, by modelling it as arising from the
space-time variation of some scalar field $\phi (\mathbf{x},t)$ whose motion
both conserved energy and momentum and created its own gravitational field
variations. The possibility that $\alpha $ varies in time has led to the
first extensive exploration of simple self-consistent theories in which $%
\alpha $ changes occur through the dynamics of some scalar field.

\section{A Simple Varying-Alpha Theory}

We consider some of the cosmological consequences of a simple theory of time
varying $\alpha .$ Such a theory was first formulated by Bekenstein \cite%
{bek} as a generalisation of Maxwell's equations but ignoring the
consequences for the gravitational field equations. We completed this theory
\ \cite{bsbm} to include the coupling to the gravitational sector and
analysed its general cosmological consequences and it is referred to as the
Bekenstein-Sandvik-Barrow-Magueijo (BSBM) theory below. Extensions to
include the weak interaction via a generalised Weinberg-Salam theory have
also been explored, \cite{km, sb}.Extensions to include the weak interaction
via a generalised Weinberg-Salam theory have also been explored, \cite{km,
sb}.

The idea that the charge on the electron, or the fine structure constant,
might vary in cosmological time was proposed in 1948 by Teller, \cite{tell},
who suggested that $\alpha \propto (\ln t)^{-1}$ was implied by Dirac's
proposal that $G\propto t^{-1}$ and the numerical coincidence that $\alpha
^{-1}\sim \ln (hc/Gm_{\text{p}}^{2})$, where $m_{\text{p }}$is the proton
mass. Later, in 1967, Gamow \cite{gam} suggested $\alpha \propto t$ as an
alternative to Dirac's time-variation of the gravitation constant, $G$, as a
solution of the large numbers coincidences problem; and in 1963 Stanyukovich
had also considered varying $\alpha $ in this context \cite{stan}. However,
any such power-law variation in the recent past was soon ruled out by other
geological evidence \cite{dyson}.

There are a number of possible theories allowing for the variation of the
fine structure constant, $\alpha $. In the simplest cases one takes $c$ and $%
\hbar $ to be constants and attributes variations in $\alpha $ to changes in 
$e$ or the permittivity of free space (see \cite{am} for a discussion of the
meaning of this choice). Thus $e_{0}\rightarrow e=e_{0}\epsilon (x^{\mu }),$
where $\epsilon $ is a dimensionless scalar field and $e_{0}$ is a constant
denoting the present value of $e$. This implies that some well established
assumptions, like charge conservation, must give way \cite{land}.
Nevertheless, local gauge invariance and causality are maintained.

Since $e$ is the electromagnetic coupling, the $\epsilon $ field couples to
the gauge field as $\epsilon A_{\mu }$ in the Lagrangian and the gauge
transformation which leaves the action invariant is $\epsilon A_{\mu
}\rightarrow \epsilon A_{\mu }+\chi _{,\mu },$ rather than the usual $A_{\mu
}\rightarrow A_{\mu }+\chi _{,\mu }.$ The gauge-invariant electromagnetic
field tensor is therefore 
\begin{equation}
F_{\mu \nu }=\frac{1}{\epsilon }\left( (\epsilon A_{\nu })_{,\mu }-(\epsilon
A_{\mu })_{,\nu }\right) ,
\end{equation}%
which reduces to the usual form when $\epsilon $ is constant. The
electromagnetic part of the action is still 
\begin{equation}
S_{em}=-\int d^{4}x\sqrt{-g}F^{\mu \nu }F_{\mu \nu }.
\end{equation}%
and the dynamics of the $\epsilon $ field are controlled by the kinetic term 
\begin{equation}
S_{\epsilon }=-\frac{1}{2}\frac{\hslash }{l^{2}}\int d^{4}x\sqrt{-g}\frac{%
\epsilon _{,\mu }\epsilon ^{,\mu }}{\epsilon ^{2}},
\end{equation}%
as in dilaton theories. Here, $l$ is the characteristic length scale of the
theory, introduced for dimensional reasons. This constant length scale gives
the scale down to which the electric field around a point charge is
accurately Coulombic. The corresponding energy scale, $\hbar c/l,$ has to
lie between a few tens of MeV and Planck scale, $\sim 10^{19}$GeV to avoid
conflict with experiment.

The field equations are 
\begin{equation}
G_{\mu \nu }=8\pi G\left( T_{\mu \nu }^{\text{matter}}+T_{\mu \nu }^{\psi
}+T_{\mu \nu }^{\text{em}}e^{-2\psi }\right) .  \label{ein}
\end{equation}%
The stress tensor of the $\psi $ field is derived from the lagrangian $%
\mathcal{L}_{\psi }=-{\frac{\omega }{2}}\partial _{\mu }\psi \partial ^{\mu
}\psi $ and the $\psi $ field obeys the equation of motion 
\begin{equation}
\square \psi =\frac{2}{\omega }e^{-2\psi }\mathcal{L}_{\text{em}}
\label{boxpsi}
\end{equation}%
where we have defined the coupling constant $\omega =(c)/l^{2}$. This
constant is of order $\sim 1$ if, as in \cite{bsbm}, the energy scale is
similar to Planck scale. It is clear that $\mathcal{L}_{\text{em}}$ vanishes
for a sea of pure radiation since then $\mathcal{L}_{\text{em}%
}=(E^{2}-B^{2})/2=0$. We therefore expect the variation in $\alpha $ to be
driven by electrostatic and magnetostatic energy-components rather than
electromagnetic radiation.

In order to make quantitative predictions we need to know how much of the
non-relativistic matter contributes to the RHS of Eqn.~(\ref{boxpsi}). This
is parametrised by $\zeta \equiv \mathcal{L}_{\text{em}}/\rho $, where $\rho 
$ is the energy density, and for baryonic matter $\mathcal{L}_{\text{em}%
}=E^{2}/2$. For protons and neutrons $\zeta _{\text{p}}$ and $\zeta _{\text{n%
}}$ can be \textit{estimated} from the electromagnetic corrections to the
nucleon mass, $0.63$ MeV and $-0.13$ MeV, respectively \cite{zal}. This
correction contains the $E^{2}/2$ contribution (always positive), but also
terms of the form $j_{\mu }a^{\mu }$ (where $j_{\mu }$ is the quarks'
current) and so cannot be used directly. Hence we take a guiding value $%
\zeta _{\text{p}}\approx \zeta _{\text{n}}\sim 10^{-4}$. Furthermore the
cosmological value of $\zeta $ (denoted $\zeta _{\text{m}}$) has to be
weighted by the fraction of matter that is non-baryonic. Hence, $\zeta _{%
\text{m}}$ depends strongly on the nature of the dark matter and can take
both positive and negative values depending on which of Coulomb-energy or
magnetostatic energy dominates the dark matter of the Universe. It could be
that $\zeta _{\text{CDM}}\approx -1$ (as for superconducting cosmic strings,
with $\mathcal{L}_{\text{em}}\approx -B^{2}/2$), or $\zeta _{\text{CDM}}\ll
1 $ (neutrinos). BBN predicts an approximate value for the baryon density of 
$\Omega _{\text{B}}\approx 0.03$ (where $\Omega _{\text{B}}$ is the density
of matter in units of the critical density $3H^{2}/8\pi G$) with a Hubble
parameter of $\ H=60$ Kms$^{\text{-1}}$ Mpc$^{\text{-1}}$, implying $\Omega
_{\text{CDM}}\approx 0.3$. Thus, depending on the nature of the dark matter, 
$\zeta _{\text{m}}$ can be virtually any number between $-1$ and $+1$. The
uncertainties in the underlying quark physics and especially the
constituents of the dark matter make it difficult to impose more certain
bounds on $\zeta _{\text{m}}$.

\subsection{The cosmological equations}

Assuming a homogeneous and isotropic Friedmann metric with expansion scale
factor $a(t)$ and curvature parameter $k$ in eqn. (\ref{ein}), we obtain the
field equations ($c\equiv 1$) 
\begin{equation}
\left( \frac{\dot{a}}{a}\right) ^{2}=\frac{8\pi G}{3}\left( \rho _{\text{m}%
}\left( 1+\zeta _{\text{m}}\exp {[-2\psi ]}\right) +\rho _{\text{r}}\exp {%
[-2\psi ]}+\frac{\omega }{2}\dot{\psi}^{2}\right) -\frac{k}{a^{2}}+\frac{%
\Lambda }{3}  \label{fried}
\end{equation}%
where $\Lambda $ is the cosmological constant. The scalar field obeys 
\begin{equation}
\ddot{\psi}+3H\dot{\psi}=-\frac{2}{\omega }\exp {[-2\psi ]}\zeta _{_{\text{m}%
}}\rho _{\text{m}},  \label{psidot}
\end{equation}%
where $H\equiv \dot{a}/a$ is the Hubble rate. We can rewrite this as

\begin{equation}
(\dot{\psi}a^{3}\dot{)}=N\exp [-2\psi ]  \label{psidot2}
\end{equation}%
where $N$ is a positive constant defined by $N=-2\zeta _{\text{m}}\rho _{%
\text{m}}a^{3}/\omega .$Note that the sign of the evolution of $\psi $ is
dependent on the sign of $\zeta _{\text{m}}$. Since the observational data
is consistent with a \emph{smaller} value of $\alpha $ in the past, we will
in this paper confine our study to \emph{negative} values of $\zeta _{\text{m%
}}$, in line with our recent discussion in Refs. \cite{bsbm}. The
conservation equations for the non-interacting radiation and matter
densities give $\rho _{\text{m}}\propto a^{-3}$ and $\rho _{\text{r}}$ $%
e^{-2\psi }\propto a^{-4},$ respectively. This theory enables the
cosmological consequences of varying $e$, to be analysed self-consistently
rather than by changing the constant value of $e$ in the standard theory to
another constant value, as in the original proposals made in response to the
large numbers coincidences. We shall consider the form of the solutions to
these equations when the universe is successively dominated by the kinetic
energy of the scalar field $\psi $, pressure-free matter, radiation,
negative spatial curvature, and positive cosmological constant$.$ Our
analytic expressions are checked by numerical solutions of (\ref{fried}) and
(\ref{psidot}). There are a number of conclusions that can be drawn from the
study of the simple BSBM models with $\zeta _{\text{m}}<0$. These models
give a good fit to the varying $\alpha $ implied by the QSO data of refs. 
\cite{webb1,webb2}. There is just a single parameter to fit and this is
given by the choice%
\begin{equation}
-\frac{\zeta _{m}}{\omega }=(2\pm 1)\times 10^{-4}  \label{om}
\end{equation}

The simple solutions predict a slow (logarithmic) time increase during the
dust era of $k=0$ Friedmann universes. The cosmological constant turns off
the time-variation of $\alpha $ at the redshift when the universe begins to
accelerate ($z\sim 0.7$) and so there is no conflict between the $\alpha $
variation seen in quasars at $z\sim 1-3.5$ and the limits on possible
variation of $\alpha $ deduced from the operation of the Oklo natural
reactor \cite{oklo} (even assuming that the cosmological variation applies
unchanged to the terrestrial environment). The reactor operated 1.8 billion
years ago at a redshift of only $z\sim 0.1$ when no significant variations
were occurring in $\alpha $. The slow logarithmic increase in $\alpha $ also
means that we would not expect to have seen any effect yet in the anisotropy
of the microwave backgrounds \cite{bat, avelino}: the value of $\alpha $ at
the last \ scattering redshift, $z=1000,$ is only 0.005\% lower than its
value today. Similarly, the essentially constant evolution of $\alpha $
predicted during the radiation era leads us to expect no measurable effects
on the products of Big Bang nucleosynthesis (BBN) because $\alpha $ was only
0.007\% smaller at BBN than it is today.

Theories in which $\alpha $ varies will in general lead to violations of the
weak equivalence principle (WEP). This is because the $\alpha $ variation is
carried by a field like $\psi $ and this couples differently to different
nuclei because they contain different numbers of electrically charged
particles (protons). The theory discussed here has the interesting
consequence of leading to a relative acceleration of order $10^{-13}$ \cite%
{bmswep} if the free coupling parameter is fixed to the value given in eq. (%
\ref{om}) by using a best fit of the theories cosmological model to the QSO
observations of refs. \cite{webb1, webb2}. Other predictions of such WEP
violations have also been made in refs. \cite{poly, zal, zald, dam}. The
observational upper bound on this parameter from direct experiment is just
an order of magnitude larger, at $10^{-12},$ and limits from the motion of
the Moon are of similar order, but space-based tests planned for the STEP
mission \cite{step} are expected to achieve a sensitivity of order $10^{-18}$
and will provide a completely independent check on theories of time-varying $%
e$ and $\alpha .$This is an exciting prospect for the future.

\subsection{The nature of the Friedmann solutions}

Let us present the predicted cosmological evolution of $\alpha $ in the BSBM
theory$,$ that we summarised above, in a little more detail. During the
radiation era the expansion scale factor of the universe increases as $%
a(t)\sim t^{1/2}$ and $\alpha $ is essentially constant in universes with an
entropy per baryon and present value of $\alpha $ like our own. It increases
in the dust era, where $a(t)\sim t^{2/3}$. The increase in $\alpha $
however, is very slow and $\alpha \sim 2N\log (t/t_{1}).$ This slow increase
continues until the expansion becomes dominated by negative curvature, $%
a(t)\sim t$, or by a cosmological vacuum energy, $a(t)\sim \exp [\Lambda
t/3] $. Thereafter $\alpha $ asymptotes rapidly to a constant. If we set the
cosmological constant equal to zero and $k=0$ then, during the dust era, $%
\alpha $ would continue to increase indefinitely. $\ $The effect of the
expansion is very significant at all times. Non-zero curvature or a
cosmological constant stops the increase in the value of $\alpha $ that
occurs during the dust-dominated era. Hence, if the spatial curvature and $%
\Lambda $ are both too\textit{\ small} it is possible for the fine structure
constant to grow too large for biologically important atoms and nuclei to
exist in the universe. There will be a time in the future when $\alpha $
reaches too large a value for life to emerge or persist. The closer a
universe is to flatness or the closer $\Lambda $ is to zero so the longer
the monotonic increase in $\alpha $ will continue, and the more likely it
becomes that life will be extinguished. Conversely, a non-zero positive $%
\Lambda $ or a non-zero negative curvature will stop the increase of $\alpha 
$ earlier and allow life to persist for longer. If life can survive into the
curvature or $\Lambda $-dominated phases of the universe's history then it
will not be threatened by the steady cosmological increase in $\alpha $
unless the universe collapses back to high density.

There have been several studies, following Carter, \cite{car} of the need
for life-supporting universes to expand close to the 'flat' Einstein de
Sitter trajectory for long periods of time. This ensures that the universe
cannot collapse back to high density before galaxies, stars, and biochemical
elements can form by gravitational instability, or expand too fast for stars
and galaxies to form by gravitational instability \cite{ch, bt}. Likewise,
it was pointed out by Barrow and Tipler, \cite{bt} that there are similar
anthropic restrictions on the magnitude of any cosmological constant, $%
\Lambda $. If it is too large in magnitude it will either precipitate
premature collapse back to high density (if $\Lambda <0$) or prevent the
gravitational condensation of any stars and galaxies (if $\Lambda >0$).
Thus, we can provide good anthropic reasons why we can expect to live in an
old universe that is neither too far from flatness nor dominated by a much
stronger cosmological constant than observed ($\left\vert \Lambda
\right\vert \leq 10\left\vert \Lambda _{\text{obs}}\right\vert $). Our
results for varying $\alpha $ suggest that there might be significant
anthropic constraints if $\Lambda $ or the spatial curvature is too small to
prevent domination before atomic structures become impossible.

\section{ Observations in Space and in the Lab}

Studies of relativistic fine structure in the absorption lines of dust
clouds around quasars by Webb et al., \cite{webb1,webb2}, have led to
widespread theoretical interest in the question of whether the fine
structure constant has varied in time.\ The quasar data analysed in refs. 
\cite{webb1,webb2} consists of three separate samples of Keck-Hires
observations which combine to give a data set of 128 objects at redshifts $%
0.5<z<3$. The many-multiplet technique finds that their absorption spectra
are consistent with a shift in the value of the fine structure constant
between these redshifts and the present of $\Delta \alpha /\alpha \equiv
\lbrack \alpha (z)-\alpha ]/\alpha =-0.57\pm 0.10\times 10^{-5},$ where $%
\alpha \equiv $ $\alpha (0)$ is the present value of the fine structure
constant \cite{webb1,webb2}. Extensive analysis has yet to find a selection
effect that can explain the sense and magnitude of the relativistic
line-shifts underpinning these deductions. A smaller study of 23 VLT-UVES
absorption systems between $0.4\leq z\leq 2.3$ by Chand \emph{et al.} \cite%
{chand} initially found $\Delta \alpha /\alpha =-0.6\pm 0.6\times 10^{-6}$
by using an approximate version of the full MM technique. However a recent
reanalysis of the same data by Murphy \emph{et al.} using the full unbiased
MM method increased the uncertainties and suggested the revised figure of $%
\Delta \alpha /\alpha =-0.64\pm 0.36\times 10^{-5}$ for the same data \cite%
{webb3}.$\ $

Any variation of $\alpha $ today can also be constrained by direct
laboratory searches. These are performed by comparing clocks based on
different atomic frequency standards over a period of months or years. Until
very recently, the most stringent constraints on the temporal variation in $%
\alpha $ arose by combining measurements of the frequencies of Sr \cite%
{blatt}, Hg+ \cite{fortier}, Yb+ \cite{peiknew}, and H \cite{fischer}
relative to Caesium: $\dot{\alpha}/\alpha =(-3.3\pm 3.0)\times 10^{-16}\,%
\mathrm{yr}^{-1}$. Cing\"{o}z \emph{et al.} also recently reported a less
stringent limit of $\dot{\alpha}/{\alpha }=-(2.7\pm 2.6)\times 10^{-15}\,%
\mathrm{yr}^{-1}$ \cite{cingoz}; however, if the systematics can be fully
understood, an ultimate sensitivity of $10^{-18}\,\mathrm{yr}^{-1}$ is
possible with their method \cite{nguyen}. If a linear variation in $\alpha $
is assumed then the Murphy \emph{et. al.} quasar measurements equate to $%
\dot{\alpha}/{\alpha }=(6.4\pm 1.4)\times 10^{-16}\,\mathrm{yr}^{-1}$. If
the variation is due to a light scalar field described by a theory like that
of BSBM \cite{bsbm}, then the rate of change in the constants is
exponentially damped during the recent dark-energy-dominated era of
accelerated expansion, and one typically predicts $\dot{\alpha}/\alpha
=1.1\pm 0.3\times 10^{-16}\,\mathrm{yr}^{-1}$ from the Murphy \emph{et al}
data, which is not ruled out by the atomic-clock constraints mentioned
above. For comparison, the Oklo natural reactor constraints, which reflect
the need for the $\mathrm{Sm}^{149}+n\rightarrow \mathrm{Sm}^{150}+\gamma $
neutron capture resonance at $97.3\,\mathrm{meV}$ to have been present $%
1.8-2\,\mathrm{Gyr}$ ($z=0.15$) ago, as first pointed out by Shlyakhter \cite%
{oklo}, are currently $\Delta \alpha /\alpha =(-0.8\pm 1.0)\times 10^{-8}$
or $(8.8\pm 0.7)\times 10^{-8}$ (because of the double-valued character of
the neutron capture cross-section with reactor temperature) and $\Delta
\alpha /\alpha >4.5\times 10^{-8}$ $(6\sigma )$ when the non-thermal neutron
spectrum is taken into account. However, there remain significant
environmental uncertainties regarding the reactor's early history and the
deductions of bounds on constants. The quoted Oklo constraints on $\alpha $
apply only when all other constants are held to be fixed. If the quark
masses to vary relative to the QCD scale, the ability of Oklo to constrain
variations in $\alpha $ is greatly reduced \cite{Flambaun07}.

Recently, Rosenband \emph{et al.} \cite{Rosenband} measured the ratio of
aluminium and mercury single-ion optical clock frequencies, $f_{\mathrm{Al+}%
}/f_{\mathrm{Hg+}}$, repeated over a period of about a year. From these
measurements, the linear rate of change in this ratio was found to be $%
(-5.3\pm 7.9)\times 10^{-17}\,\mathrm{yr}^{-1}$. These measurements provides
the strongest limit yet on any temporal drift in the value of $\alpha $: $%
\dot{\alpha}/\alpha =(-1.6\pm 2.3)\times 10^{-17}\,\mathrm{yr}^{-1}$. This
limit is strong enough to strongly rule out theoretical explanations of the
change in $\alpha $ reported by Webb \emph{et al.} \cite{webb1, webb2} in
terms of the slow variation of an effectively massless scalar field, even
allowing for the damping by cosmological acceleration, unless there is a
significant new physical effect that slows the locally observed effects of
changing $\alpha $ on cosmological scales (for a detailed analysis of
global-local coupling of variations in constants see Refs. \cite{shaw}).

It has been noted that if the `constants' such as $\alpha $ or $\mu $ can
vary, then in addition to a slow temporal drift one would also expect to see
an annual modulation in their values. In many theories, the Sun perturbs the
values of the constants by a factor roughly proportional to the Sun's
Newtonian gravitational potential (the contribution from the Earth's
gravitational potential is about 14 times smaller than that of the Sun's at
the Earth's surface). Hence the `constants' depend on the distance from the
Sun. Since the Earth's orbit around the Sun has a small ellipticity, the
distance, $r$, between the Earth and Sun fluctuates annually, reaching a
maximum at aphelion around the beginning of July and a minimum at perihelion
in early January. It was shown in Ref. \cite{shawb} that in many varying
constant models, the values of the constants measured here on Earth, would
oscillate in a similar seasonal manner. Moreover, in many cases, this
seasonal fluctuation is predicted to dominate over any linear temporal drift 
\cite{shawb}.

Specifically, let us suppose that the Sun creates a distance-dependent
perturbation to the measured value of a coupling constant, $\mathcal{C}$, of
amplitude $\delta \ln \mathcal{C}=C(r)$. If this coupling constant is
measured on the surface of another body (e.g. the Earth) which orbits the
first body along an elliptical path with semi-major axis $a$, period $T_{%
\mathrm{p}}$, and eccentricity $e\ll 1$, then to leading order in $e$, the
annual fluctuation in $\mathcal{C}$, $\delta \mathcal{C}_{\mathrm{annual}}$
will be given by 
\begin{equation}
\frac{\delta \mathcal{C}_{\mathrm{annual}}}{\mathcal{C}}=-c_{\mathcal{C}%
}\cos \left( \frac{2\pi t}{T_{\mathrm{p}}}\right) +O(e^{2}),  \label{VarForm}
\end{equation}%
where $c_{\mathcal{C}}\equiv e\,aC^{\prime }(a)$, $C^{\prime
}(a)=dC(r)/dr|_{r=a}$ and $t=nT_{\mathrm{p}}$, for any integer $n$,
corresponds to the moment of closest approach (perihelion). In the case of
the Earth moving around the Sun, over a period of 6 months from perihelion
to aphelion one would therefore measure a change in the constant $\mathcal{C}
$ equal to $2c_{\mathcal{C}}$. Using $\delta \ln (f_{\mathrm{Al+}}/f_{%
\mathrm{Hg+}})=(3.19+0.008)\delta \alpha /\alpha $, \cite{Rosenband}, a
maximum likelihood fit to the data gives $c_{\alpha }=ea\delta \alpha
^{\prime }(a)=\left( -0.89\pm 0.84\right) \times 10^{-17}$, where $%
a=149,597,887.5\,\mathrm{km}$ is the semi-major axis of the Earth's orbit,
and $\delta \alpha (r)$ is the perturbation in $\alpha $ due to the Sun's
gravitational field. Assume that over solar system scales, the values of the
scalar fields on which values of the 'constants' depend, vary with the local
gravitational potential. Hence, we have $\delta \alpha (r)/\alpha =k_{\alpha
}\Delta U_{\odot }(r),$ where $k_{\alpha \text{ }}$is a theory-dependent
multiplier, $\Delta U_{\odot }$ is the change in the gravitational potential
of the Sun: $U_{\odot }(r)=-GM_{\odot }/r$, and so $ea\Delta U_{\odot
}^{\prime }(a)=eGM_{\odot }/a=1.65\times 10^{-10}$. Hence, we find: 
\begin{equation}
k_{\alpha }=(-5.4\pm 5.1)\times 10^{-8}.  \label{kalpha}
\end{equation}%
The frequency shifts measured by Rosenband \emph{et al.} \cite{Rosenband}
were not sensitive to changes in the electron-proton mass ratio: $\mu =m_{%
\mathrm{e}}/m_{\mathrm{p}}$. Measurements of optical transition frequencies
relative to Cs, Refs. \cite{blatt, fortier, peiknew, fischer}, are sensitive
to both $\mu $ and $\alpha $. H-maser atomic clocks \cite{Ashby} are also
sensitive to variations in the light quark to proton mass ratio: $q=m_{%
\mathrm{q}}/m_{\mathrm{p}}$. We can use all these observations if we define
two more gravitational coupling multipliers, $k_{\mu }$ and $k_{\mathrm{q}}$%
, by $\delta \mu /\mu =k_{\mu }\Delta U_{\odot }$, and $\delta q/q=k_{%
\mathrm{q}}\Delta U_{\odot }$. Refs. \cite{blatt, fortier, Ashby} give $%
k_{\alpha }+0.36k_{\mu }=(-2.1\pm 3.2)\times 10^{-6}$, $k_{\alpha
}+0.17k_{\mu }=(3.5\pm 6.0)\times 10^{-7}$, and $k_{\alpha }+0.13k_{\mathrm{q%
}}=(1\pm 17)\times 10^{-7}$ respectively. We also performed a bootstrap
seasonal fluctuation fit (with $10^{5}$ resamplings) to the $\mathrm{Yb}^{+}$
frequency measurements of Peik \emph{et al.} \cite{peiknew} giving $%
k_{\alpha }+0.51k_{\mu }=(7.1\pm 3.4)\times 10^{-6}$ \cite{bshaw}. Combining
these bounds with Eq. (\ref{kalpha}) gives $k_{\mu }=(3.9\pm 3.1)\times
10^{-6}$, $k_{\mathrm{q}}=(0.1\pm 1.4)\times 10^{-5}$. Recently, Blatt \emph{%
et al.} \cite{blatt} combined data from measurements of H-maser \cite{Ashby}
and optical atomic clocks \cite{blatt, fortier}, to bound the multipliers, $%
k_{\alpha }$, $k_{\mu }$ and $k_{q}$, finding $k_{\alpha }=(2.5\pm
3.1)\times 10^{-6}.$The constraint on $k_{\alpha }$ derived in this paper
from the data of Rosenband \emph{et al.} \cite{Rosenband} therefore
represents an improvement by about two orders of magnitude over the previous
best bound. This improved bound on $k_{\alpha }$ combined with data found by
Peik \emph{et al.} \cite{peiknew} has also produced an order of magnitude
improvement in the determination of $k_{\mu }$ and a slight improvement in
the constraint on $k_{q}$.

Seasonal fluctuations are predicted by a varying constant theory because the
scalar field which drives the variation in the constant couples to normal
matter. The presence of the Sun therefore induces gradients in scalar fields
and associated varying 'constants', and it is essentially these gradients
that are detectable as seasonal variables. As mentioned earlier, gradients
in a scalar field which couples to normal matter result in new or 'fifth'
forces with pseudo-gravitational effects. In the case of varying $\alpha $
and $\mu $ theories, these forces are almost always composition dependent,
which would violate the universality of free-fall and hence the weak
equivalence principle (WEP). The magnitude of any composition-dependent
fifth force toward the Sun is currently constrained to be no stronger than $%
10^{-12}-10^{-13}$ times than the gravitational force \cite{WEP}. In the
context of a given theory the constraints from WEP tests indirectly bound $%
k_{\alpha }$. Indeed, they often provide the tightest constraints on $%
k_{\alpha }$ \cite{bshaw, dent, bmag2}.

A recent thorough analysis of the WEP violation constraints on $k_{\alpha }$ 
\cite{dent,dentprivate} found $k_{\alpha }=(0.3\pm 1.7)\times 10^{-9}$, with
a similar constraint on $k_{q}$. It must be noted, however, that this result
is still subject to theoretical uncertainty, especially regarding the
dependence of nuclear properties on quark masses. For instance, it was also
noted in Ref. \cite{dent} that if certain (fairly reasonable) assumptions
about nuclear structure are dropped, the 1$\sigma $ error bars on $k_{\alpha
}$ increase by about an order of magnitude to: $\pm 1.4\times 10^{-8}$.
Despite these uncertainties, for many theories of varying $\alpha $, WEP
violation constraints from laboratory experiments or lunar laser ranging 
\cite{nord} still provide the strongest, albeit indirect, bound on $%
k_{\alpha }$.

\section{Conclusions}

We have described some of the history of the study of varying constants in
physics. This area of research has been reinvigorated by a significant body
of observational data, drawn from quasar absorption spectra, which is
consistent with a change in the value of the fine structure constant, $%
\alpha $, over of a few parts in a million over 10 billion years. So far,
these data have neither been reliably confirmed nor contradicted by other
observational studies and this confrontation is keenly awaited. We described
how a simple self-consistent theory of varying $\alpha $ developed by
Sandvik, Barrow and Magueijo can be constructed and the clear pattern of
variation that it predicts in the universe: no variation of $\alpha $ during
the radiation era and a logarithmic time increase during the
cold-dark-matter era, followed by a resumption of no time variation in $%
\alpha $ after the universe begins to accelerate during the dark-energy era.
A fit of these simple models to the observational data fixes the one free
parameter defining the theory and predicts a violation of the weak
equivalence principle at a $10^{-13}$ level that is easily detectable from
space. We also described exciting new developments in the laboratory search
for varying $\alpha $. These experiments are for the first time achieving
the sensitivities of the indirect astronomical bounds and in the next few
years we may be able to draw some strong conclusions about varying constants
from the confluence of laboratory experiments and large new astronomical
data sets.

\bigskip 

\textbf{Acknowledgements }I am very grateful to Mariusz Dabrowski for his
help and hospitality during the Grassmann Bicentenial Conference in Szczecin.

\end{document}